\documentclass[fleqn,usenatbib]{mnras}

\usepackage[T1]{fontenc}

\DeclareRobustCommand{\VAN}[3]{#2}
\let\VANthebibliography\thebibliography
\def\thebibliography{\DeclareRobustCommand{\VAN}[3]{##3}\VANthebibliography}

\usepackage{graphicx}	% Including figure files
\usepackage{amsmath}	% Advanced maths commands
\usepackage{amssymb}	% Extra maths symbols
\usepackage{newtxtext,newtxmath}

\title[Gas clumping in the MACSIS simulations]{Gas clumping and its effect on hydrostatic bias in the MACSIS simulations}

\author[I. Towler et al.]{
Imogen Towler,$^{1}$\thanks{E-mail: imogen.towler@manchester.ac.uk}
Scott T. Kay $^{1}$
 and Edoardo Altamura $^{1}$
\\
$^{1}$Jodrell Bank Centre for Astrophysics, School of Physics and Astronomy, The University of Manchester, Manchester M13 9PL, UK\\
}

\date{Accepted XXX. Received YYY; in original form ZZZ}

\pubyear{2022}

\begin{document}
\label{firstpage}
\pagerange{\pageref{firstpage}--\pageref{lastpage}}
\maketitle

\begin{abstract}
We use the MACSIS hydrodynamical simulations to estimate the extent of gas clumping in the intracluster medium of massive galaxy clusters and how it affects the hydrostatic mass bias. By comparing the clumping to the azimuthal scatter in the emission measure, an observational proxy, we find that they both increase with radius and are larger in higher-mass and dynamically perturbed systems. Similar trends are also seen for the azimuthal temperature scatter and non-thermal pressure fraction, both of which correlate with density fluctuations, with these values also increasing with redshift. However, in agreement with recent work, we find only a weak correlation between the clumping, or its proxies, and the hydrostatic mass bias. To reduce the effect of clumping in the projected profiles, we compute the azimuthal median following recent observational studies, and find this reduces the scatter in the bias. We also attempt to correct the cluster masses by using a non-thermal pressure term and find over-corrected mass estimates ($1-b=0.86$ to $1-b=1.15$) from 3D gas profiles but improved mass estimates ($1-b=0.75$ to $1-b=0.85$) from projected gas profiles, with the caveat of systematically increased scatter. We conclude that the cluster-averaged mass bias is minimised from applying a non-thermal pressure correction ($1-b=0.85$) with more modest reductions from selecting clusters that have low clumping ($1-b=0.79$) or are dynamically relaxed ($1-b=0.80$). However, the latter selection is most effective at minimising the scatter for individual objects. Such results can be tested with next generation X-ray missions equipped with high-resolution spectrometers such as Athena.
\end{abstract}


%%%%%%%%%%%%%%%%%%%%%%%%%%%%%%%%%%%%%%%%%%%%%%%%%%

%%%%%%%%%%%%%%%%% BODY OF PAPER %%%%%%%%%%%%%%%%%%

\section{Introduction}
\label{sec:intro}

Galaxy clusters originate from the largest fluctuations in the early universe and today, they are the the largest virially relaxed structures. Their constituent galaxies evolve with the cluster through a series of processes such as star formation, radiative cooling and AGN feedback. Cluster studies therefore allow the study of astrophysical baryonic processes and their interplay with cosmology.

The matter power spectrum of primordial fluctuations and their growth can be studied using the distribution of galaxy clusters and their masses and redshifts. This results in the ability to place constraints on the fundamental cosmological constants of the Universe \citep[see][for a review]{Borgani2011}. However, such constraints require highly accurate measurements of cluster mass through methods that result in a low mass bias and scatter or using methods to correct the mass estimates. The discrepancies between cluster mass estimates and their true masses have been proposed to be the cause of at least some of the difference between the normalisation of the power spectrum of density fluctuations, $\sigma_8$, when comparing the value determined from cluster number counts to using anisotropies in the cosmic microwave background \citep{Planck2014xx}. \citet{Planck2014i} proposed that this discrepancy could be caused by an underestimate in the hydrostatic masses determined for the galaxy cluster sample.

To measure the hydrostatic mass of a cluster, X-ray observations are used to measure the density (via their surface brightness) and temperature (via spectroscopic measurements) profiles. These are combined under the assumption that the cluster is spherically symmetric and in hydrostatic equilibrium. However, in practice, the intracluster medium (hereafter ICM) is not evenly distributed, deviating from spherical symmetry. In addition, the cluster gas will have residual motions \citep[e.g.][]{Rasia2004AClusters, Lau2009ResidualClusters, Vazza2009TurbulentRefinement, Biffi2016OnClusters}, breaking the hydrostatic equilibrium. This introduces a bias to the hydrostatic mass, the extent of which is currently not fully agreed upon.

Observations can estimate the mass bias by comparing the X-ray hydrostatic mass to the weak lensing mass \citep[although weak lensing itself is thought to be biased by around 5-10 per cent due to projection effects and asphericity, e.g.][]{Becker2011OnReconstructions, Bahe2012MockConcentration, Henson2017}. The hydrostatic mass bias\footnote{We define the bias as $1-b = M_{\rm{HSE}} / M_{\rm{true}}$, where $M_{\rm{HSE}}$ is the hydrostatic mass and $M_{\rm{true}}$ is the true mass}, $b$, has previously been measured to be $b < 0.1$ \citep[e.g.][]{Applegate2014WeighingMasses, Israel2014TheProgramme, Smith2016LoCuSS:Clusters}, contradicting the larger bias, $b=0.2-0.3$, found by \citet{Hoekstra2015TheMasses, Battaglia2016Weak-lensingSurvey, Miyatake2019Weak-lensingSurvey}. \citet{Umetsu2022Line-of-sight370} found a bias as high as $b=0.44-0.49$ for a particularly perturbed galaxy cluster.

Simulations allow the direct comparison of the hydrostatic mass of a cluster to its true mass. When using mass-weighted gas profiles, the bias has been found to be approximately  $b=0.1-0.2$ \citep[e.g.][]{Rasia2006SystematicsEstimators, Lau2009ResidualClusters, Roncarelli2013, Biffi2016OnClusters, Henson2017, Angelinelli2020TurbulentMedium, Pearce2020, Ansarifard2020, Barnes2021CharacterizingMock-X, Gianfagna2021ExploringSample}. Mock observations can be produced using simulation data and then used to estimate additional observational effects on the mass bias.
Works such as \citet{Lau2009ResidualClusters, Rasia2012, LeBrun2014} found little difference between measuring the mass bias through mass-weighted and spectroscopic profiles. However, other studies \citep[e.g.][]{Henson2017, Barnes2017a, Barnes2021CharacterizingMock-X} found an increase in the bias using spectroscopic profiles.

The dependence of bias on cluster properties, such as cluster morphology or mass, has also been investigated in simulations, with some contradictory results. For example, works such as \citet{Henson2017, Pearce2020} and \citet{Barnes2021CharacterizingMock-X} have found a mass dependence in the bias when using spectroscopic profiles derived from mock X-ray images. In addition, \citet{Barnes2017a} found no difference in the average mass bias of relaxed and perturbed clusters in the C-EAGLE cluster simulations, only a reduced scatter in relaxed clusters, whereas \citet{Ansarifard2020} found that more relaxed clusters have a lower bias.

This work aims to look at the effects on the mass bias caused by the inhomogeneity of the ICM. Cluster gas in simulations is not smoothly distributed; it contains small clumps of gas and larger scale density fluctuations \citep{Roncarelli2006SimulatedProfiles, Nagai2011GASCLUSTERS, Vazza2011, Churazov2012X-rayCluster, Morandi2013Non-parametricClusters}. The brightness of X-ray images is proportional to the gas density squared, and so recovered gas density profiles from these observations are likely to be biased and higher than otherwise expected \citep{Nagai2011GASCLUSTERS, Vazza2013PropertiesMedium, Zhuravleva2013, Roncarelli2013, Planelles2017}. Density fluctuations in the ICM are quantified using the clumping parameter, defined as
\begin{equation}
    \mathscr{C} = \frac{ \left< \rho^{2} \right> }{\left< \rho \right> ^2},
    \label{eq:clumping_def}
\end{equation}
where $\rho$ is the gas density and the angled brackets denote the mean for the region in question. For a perfectly uniform medium, we would calculate $\mathscr{C} = 1$, and for all other cases $\mathscr{C} > 1$.

Past works, such as \citet{Roncarelli2013, Eckert2015} and \citet{Ansarifard2020}, have studied the effects of the clumping by relating it to quantities that X-ray observers could measure, in particular the azimuthal scatter, $\sigma_{A}$, which splits annuli used to take profile measurements into azimuthal bins and calculates the scatter in the emission measure relative to the median. \citet{Roncarelli2013} found a strong correlation between the azimuthal scatter and the residual clumping and used this relationship to estimate the clumping using observables and correct the hydrostatic mass. They succeeded in reducing the bias in the hydrostatic mass but the large scatter in the bias remained. However, \citet{Ansarifard2020} found no correlation between the clumping and the bias and deduced that it cannot be used to reduce the mass bias. Instead they reduced the bias and its scatter by including additional corrections using the slope of the gas density and pressure profiles. Alternatively, \citet{Zhuravleva2013} and \citet{Eckert2015} found that using the azimuthal median instead of the mean density, temperature and pressure profiles reduced the effects of inhomogeneities in the profiles and removed the need to account for clumping. 

In this paper, we investigate the impact of gas inhomogeneities in the MACSIS simulation \citep{Barnes2017}, a hot and massive sample of galaxy clusters. The level of gas inhomogeneities is calculated via the clumping parameter and compared to observational proxies. In addition to the effect on extracted density profiles, gas inhomogeneities in the temperature and velocity are also investigated. The hydrostatic masses are calculated for the MACSIS sample using both the 3D true gas profiles and using projected, 2D gas profiles, with relevant weightings to mimic spectroscopic profiles. We also present the effect of projection on the mass estimate and how mass corrections using the velocity dispersion affect 2D and 3D mass estimates.  

The rest of the paper is structured as follows. Section \ref{sec:sims} briefly describes the MACSIS simulated cluster sample and then outlines how maps and profiles have been calculated from this data set and how the clusters' dynamical states have been classified. The effects of gas inhomogeneities on gas profiles, and how we have minimised the effect of this, are described in Section \ref{sec:inhomogeneities}. Section \ref{sec:mass_bias} presents the results of the estimated hydrostatic masses of the clusters, how this can be corrected and whether this correlates with any observable or proxy for ICM inhomogeneities. Finally, our results are summarised and conclusions drawn in Section \ref{sec:conclusion}.

\section{MACSIS Simulations}
\label{sec:sims}

This work uses clusters from the MACSIS cosmological zoom simulations \citep{Barnes2017}. The parent simulation used dark matter only in a periodic box with a comoving side length of 3.2 Gpc, 2520$^{3}$ particles and a particle mass of $5.43\times 10^{10} h^{-1} \mathrm{M}_{\odot}$. The cosmological parameters used are from the Planck 2013 data release \citep{Planck2014i}; these are $h=0.6777, \Omega_{\mathrm{m}} = 0.307, \Omega_{\Lambda}=0.693, \Omega_{\mathrm{b}} = 0.04825, \sigma_{8} = 0.8288$.

From this parent simulation, clusters were chosen using their \textit{Friends-of-Friends} (FoF) mass  with a linking length of $b=0.2$ \citep{DEFW1985}.
The clusters were spread over a series of mass bins with width $\Delta \log_{10} M_{\mathrm{FoF}} = 0.2$ between $10^{15} \leq M_{\mathrm{FoF}}/\rm{M}_{\odot} \leq 10^{16}$. All clusters were selected from bins containing fewer than 100 clusters, therefore all clusters with $M_{\rm{FoF}} > 10^{15.6} \rm{M}_{\odot}$ were selected. Below this mass, the bins were further divided into ten smaller bins with $\Delta \log_{10} M_{\rm{FoF}} = 0.02$. From each of these bins, ten clusters were selected at random, giving 390 clusters in total.

These objects were re-simulated with gas and star particles and the resolution was increased using the zoom simulation technique \citep{Katz1993, Tormen1997}. These simulations were run using the same hydrodynamics, resolution (particle masses $m_{\rm{DM}}=4.4\times10^9 h^{-1}\rm{M}_{\odot}$ and $m_{\rm{gas}}=8.0\times10^8 h^{-1}\rm{M}_{\odot}$) and sub-grid physics models as BAHAMAS \citep{McCarthy2017} using GADGET-3, an updated version of the GADGET-2 code \citep{Spring2005}, with further modifications made for the OWLS project \citep{Schaye2010}. From these 390 clusters, thirteen were found to have gas fractions an order of magnitude lower than expected and so we only use a maximum sample of 377 objects throughout this work. We study the properties of the remaining objects at two redshifts: $z=0$ and $z=0.5$, approximately spanning the redshift range of the recent CHEX-MATE sample \citep[][]{CHEX-MATECollaboration2021TheOverview}.

\subsection{Map and profile definitions}
\label{sec:profs}
In this work, we study properties calculated from projected 2D and unprojected 3D data. The 3D profiles are obtained by centering a series of spherical shells on the SUBFIND \citep{SWTK2001} determined potential minimum and calculating the relevant property in each shell. The 3D profiles are used to represent the theoretical case and so these use all the gas particles in the simulation within the relevant radii, with the exception of clumping (see below). The MACSIS simulation has produced a very massive and hot sample of clusters and so using a temperature cut to calculate the 3D profiles would have a negligible effect as the gas below $kT \approx 0.3$ keV accounts for on average only one per cent of the total gas mass within $R_{500}$.

Projected profiles are calculated using the same method but with cylindrical bins and a total depth of 3$R_{500}$ \footnote{We define $R_{500}$ as the radius of a sphere such that the average density within it is 500 times the critical density of the universe. Likewise, $M_{500}$ is defined as the total mass within $R_{500}$.} along the line of sight. The clusters were projected down three perpendicular axes, resulting in three 2D profiles for each cluster. These profiles are designed to mimic the effects of an X-ray observation, so they account for projection and weightings that arise from using spectroscopic profiles, as if they were retrieved from X-ray images. Only very hot gas in the ICM emits X-rays and so a temperature cut is implemented such that only gas particles with a temperature above $kT \approx 0.3$ keV are used to obtain two-dimensional profiles, excluding the cooler and denser gas particles. Furthermore, we also split each radial bin into 12 azimuthal bins and take the mean or median (see Section \ref{sec:radial} below). 

To calculate hydrostatic masses, models for the temperature, density and non-thermal pressure fraction (following \citealt{Vikhlinin2006} for the temperature and density, and \citealt{Nelson2014} for the non-thermal pressure fraction) are fitted to the respective profiles over the radial range 0.5-1.5$R_{500}$. These models describe 3D profiles and so the models themselves are projected when used to fit the 2D data, following
\begin{equation}
    f(R) = \int _{R} ^{\infty} \frac{f(r) r \mathrm{d} r}{\sqrt{r^{2} - R^{2}}},
    \label{eq:projection}
\end{equation}
where $f(r)$ is the 3D model value at a spherical radius $r$ and $f(R)$ is the projected 2D model at a projected radius $R$. The non-thermal pressure fraction could not be projected in the same way due to weighting effects on the profiles and so, in 2D, the velocity dispersion was used instead. A combination of this model and the temperature model allowed the calculation of a non-thermal pressure model for the 2D data, as explained in Section \ref{sec:vel_alpha} below. For some of the projected profiles, the fitting did not yield well matching profiles, due to particularly noisy profiles in lower mass clusters. Therefore, the hydrostatic mass was not calculated and they have been excluded from the sample presented in Section \ref{sec:mass_bias} looking at the effects of mass bias. Specifically, 107/1131 profiles (9 per cent) were excluded from measuring the standard hydrostatic mass bias and a further four were excluded from investigation of the bias using non-thermal pressure fractions in Sections \ref{sec:non-therm} and \ref{sec:clump_bias}.

The fitting parameters from the projected profiles can then be used with the unprojected models to produce model 3D profiles, which are used for the mass estimates. We describe the procedure for each profile in the sections below.

\subsubsection{Gas density and emission measure}

To obtain 3D density profiles, a series of spherical bins were applied around the centre of a cluster. The density was calculated using the total volume of each bin and the total mass of the particles within it.

For the 2D case, emission measure ($EM$) profiles were used as mock observational data of the clusters. The emission measure is related to the total surface brightness ($SB$) in a soft X-ray band via
\begin{equation}
    SB(R) = \frac{EM(R) \Lambda(T,z)}{4\pi (1+z)^3},
    \label{eq:surface_brightness}
\end{equation}
where $\Lambda(T,z)$ is the soft X-ray cooling function, which depends on the gas temperature ($T$) and redshift ($z$), and the emission measure can be calculated using the electron and hydrogen number density ($n_{\rm{e}}$ and $n_{\rm{H}}$ respectively), projected along the line of sight
\begin{equation}
    EM = \int n_{\rm{e}} n_{\rm{H}} \, \mathrm{d}l.
    \label{eq:emission_measure_integration}
\end{equation}

The emission measure is directly related to the density of the gas so we can use emission measure profiles as a proxy for density profiles. We calculate the emission measure profile by using a series of cylindrical bins placed around the centre of a cluster. The axis down the length of the cylinder is aligned with the simulation axis that has been chosen as the line of sight for that profile. The emission measure for each radius can then be calculated by summing the contributions from the $N_{\rm{bin}}$ hot gas particles within the cylindrical bin following,
\begin{equation}
    EM(R) = \frac{X_{\mathrm{H}}}{\mu_e m_{\mathrm{H}}^2 A_{\mathrm{bin}}} \sum^{N_{\rm{bin}}}_{i=1}\rho_{i} m_{i},
\end{equation}
where $m_{\rm{H}}$ is the mass of a proton, $A_{\mathrm{bin}}$ is the cross-sectional area of the cylindrical bin, $X_{\rm{H}} = 0.76$ is the hydrogen mass fraction and $\mu_e = 1.14$ is the mean molecular weight per free electron, assuming a fully ionised primordial gas of hydrogen and helium.

Once the profiles have been calculated, they were fitted using the Vikhlinin model for gas density,
\begin{equation}
    n_{\rm{e}}n_{\rm{H}} = n_{0}^2 \frac{ \left( r/r_{\rm{c}} \right)^{-\alpha} }{ \left(1 + r^2/r_{\rm{c}}^2 \right)^{3 \beta - \alpha/2}} \frac{1}{\left(1 + r^{\gamma} / r_{\rm{s}}^{\gamma} \right) ^{\epsilon / \gamma}},
\end{equation}
where $\{ n_0$, $r_{\rm{c}}$, $\alpha$, $\beta$, $\epsilon$, $r_{\rm{s}}\}$ are free parameters used for the fitting and we assume $\gamma=3$ following \citet{Vikhlinin2006}. This model was projected following Equation \ref{eq:projection} to fit the emission measure profiles from the simulation data.

\subsubsection{Temperature}

To calculate the 3D temperature profile, a series of spherical bins are placed around the centre of a cluster and a weighted temperature is calculated for each bin, following
\begin{equation}
    T(r) = \frac{\Sigma_i W_i T_i}{\Sigma_i W_i},
\end{equation}
where, in the 3D case, we use a mass weighted temperature profile and so the weighting, $W_{i}=m_{i}$, is the mass of the $i^{\rm{th}}$ gas particle.

For the projected 2D profiles, we measure the temperature again using cylindrical bins with a total depth of 3$R_{500}$. To mimic observations, the spectroscopic-like (SL) temperature, $T_{\rm{SL}}$, is used following \citet{Mazzotta2004}, where $W_i = m_i \rho_i T_i^{-3/4}$. \citet{Rasia2005} and \citet{Roncarelli2018MeasuringObservations} have shown that temperature profiles extracted from synthetic observations of hot clusters match the spectroscopic-like temperature and give lower values than the mass-weighted or emission-weighted temperatures.

Both the 2D and 3D profiles were fitted using the model from \citet{Vikhlinin2006},
\begin{equation}
    T(r) = T_{0} \frac{ \left( r/r_{\rm{cool}} \right)^{a_{\rm{cool}}} + T_{\rm{min}}/T_0 }{ \left( r/r_{\rm{cool}} \right)^{a_{\rm{cool}}} + 1} \frac{ \left(r/r_{\rm{t}} \right)^{-A}}{ \left( 1 + r^{B}/r_{t}^{B} \right)^{C/B}},
\end{equation}
where $\{T_{0}$, $r_{\rm{cool}}$, $a_{\rm{cool}}$, $T_{\rm{min}}$, $r_{\rm{t}}$, $A$, $B$, $C\}$ are free parameters. Again, the model is projected to fit the SL temperature following Equation \ref{eq:projection}.

\begin{figure*}
    \centering
    \includegraphics[trim={1.5cm 7.5cm 1.5cm 2cm}, clip]{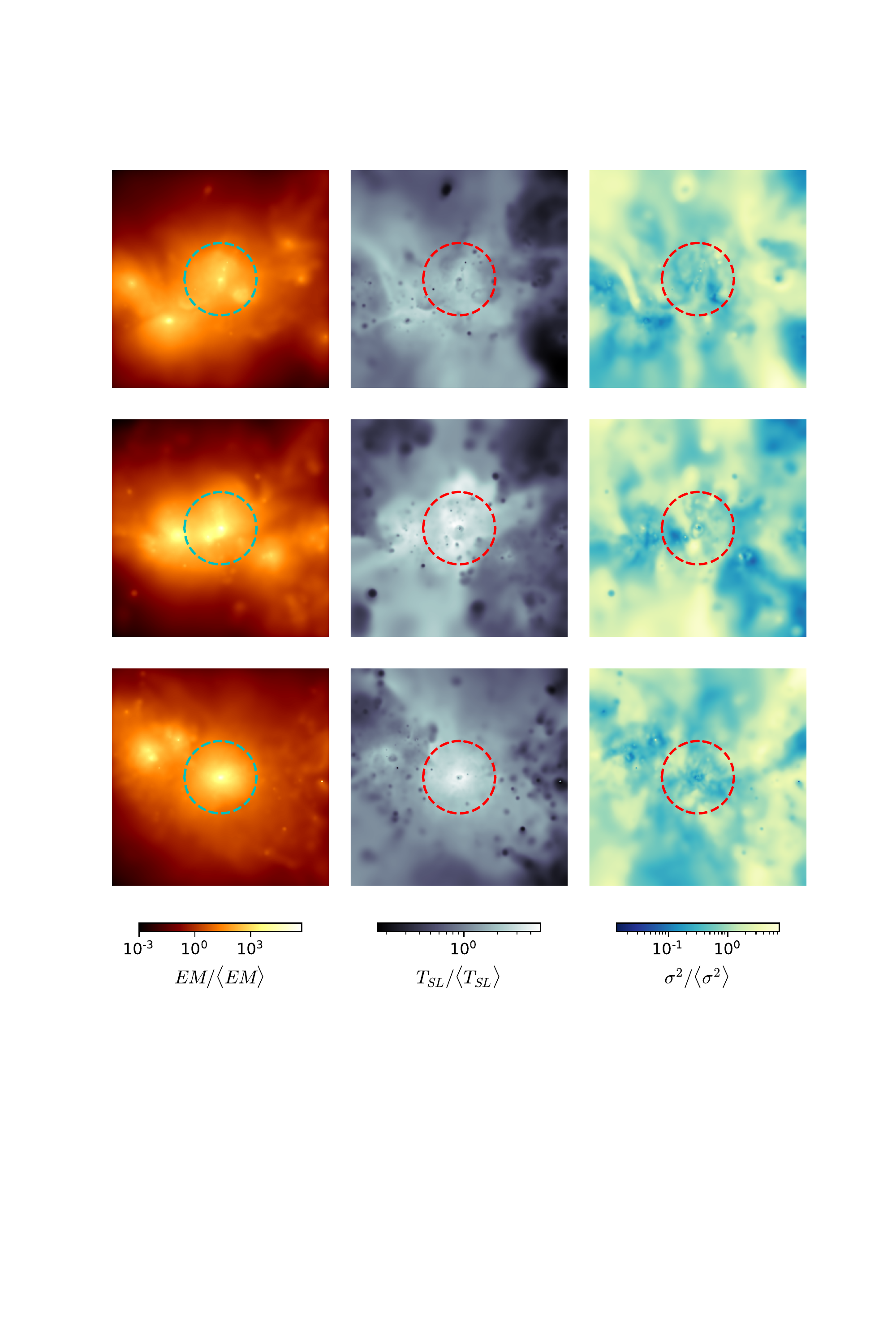}
    \caption{Emission measure (left column), spectroscopic-like temperature (middle) and velocity dispersion (right) maps of three MACSIS clusters. The maps have been normalised by their respective mean pixel value to reduce the range of values between the maps. The top row shows a perturbed lower mass (7.0$\times 10^{14} \rm{M_{\odot}}$) cluster at $z=0$. The middle and bottom rows both show the same cluster at two epochs, with the middle row at $z=0.5$ and the bottom at $z=0$; at $z=0$ it is relaxed, but at $z=0.5$ is perturbed. It is a high mass cluster, at $z=0$, $M_{500} =$ 1.4$\times 10^{15} \rm{M}_{\odot}$. The central circles in the maps show a radius of $R_{500}$, with a total side length of 6 $R_{500}$ for each map.}
    \label{fig:halos_103_51_51}
\end{figure*}

\subsubsection{Velocity dispersion and non-thermal pressure fraction}
\label{sec:vel_alpha}

Clusters are often assumed to be relaxed systems, however turbulence and bulk motions within the ICM can contribute to the amount of non-thermal pressure within a cluster. While it is not currently possible to measure the peculiar motions of gas within clusters extensively, future X-ray telescopes such as \textit{Athena} will soon be able to do so \citep{Nandra2013TheMission}.

We estimate the non-thermal pressure, $P_{\mathrm{nth}}$, via
\begin{equation}
    P_{\mathrm{nth}} = \frac{1}{3} \rho_{\mathrm{gas}} \sigma_{\mathrm{3D}}^2,
    \label{eq:non-therm_pressure}
\end{equation}
using the gas density and the three-dimensional velocity dispersion, $\sigma_{\mathrm{3D}}$.

The 1D velocity dispersion, $\sigma_{j}$, is determined from the velocity components along a chosen axis ($j \in \{x,y,z\}$ or $j \in \{r, t\}$, where $r$ and $t$ are the radial and tangential components respectively), defined as
\begin{equation}
    \sigma_{j}^{2} = \frac{\sum_{i} W_{i} \left( v_{i,j} - \bar{v}_{j} \right)^{2}}{\sum_{i} W_{i}}
\end{equation}
where $v_{i,j}$ is the velocity component of the $i^{\rm{th}}$ simulated particle down axis $j$, $W_{i}$ is the relevant weighting and $\bar{v}_{j}$ is the average velocity in the respective bin. $\bar{v}_{j}$ is subtracted from each particle's velocity, $v_{i,j}$,
to account for the cluster's bulk velocity \footnote{In the 3D case, this is taken as the mean mass-weighted velocity in the spherical shell. In 2D, we subtract the bulk velocity along the lines of sight within each of the 12 azimuthal bins and then take the median value of $\sigma_{\rm{los}}^2$.}. For the 3D profiles in this work, we use $W_{i} = m_{i}$ and the 1D velocity dispersion is calculated for three perpendicular axes and then summed in quadrature to obtain the 3D velocity dispersion. 
For the projected profiles, we mimic observations and only consider the velocity component along the line of sight (los). We use an emission weighting, $W_{i} = \rho_{i} m_{i}$, following \citet{Roncarelli2018MeasuringObservations} who found that the emission weighting reproduces the spectroscopic velocity and velocity dispersion well in the context of future Athena X-IFU observations. To calculate the 3D velocity dispersion, we assume an isotropic velocity distribution and use $\sigma_{3\mathrm{D}}^2 = 3 \sigma_{\mathrm{los}}^2$. 

In this work, when estimating a cluster's total mass, we use the non-thermal pressure fraction instead of the velocity dispersion directly, defined as
\begin{equation}
    \alpha = \frac{P_{\mathrm{nth}}}{P_{\mathrm{tot}}} = \left[ 1 + \frac{3 k_{\rm{B}} T}{\mu m_{\rm{H}} \sigma_{\rm{3D}}^2} \right]^{-1},
    \label{eq:alpha}
\end{equation}
where $\mu=0.59$ is the mean molecular weight. Unprojected profiles use the 3D velocity dispersion with the mass weighted temperature; these are fitted following \citet{Nelson2014}, 
\begin{equation}
    \alpha(r) = 1 - A \left( 1 + \exp \left[- \left( \frac{r/r_{500}}{B} \right) ^{\rm{C}} \right] \right),
\end{equation}
where $\{A,B, C\}$ are free parameters to be determined in the fitting.

In contrast, the projected non-thermal pressure fraction profiles use the projected spectroscopic-like temperature profile and the velocity dispersion measured along the line of sight. The weightings used to obtain the profiles are required to normalise the deprojected model. For the non-thermal pressure fraction, both the velocity dispersion and SL temperature have separate weightings which cannot be used in the deprojection simultaneously. Instead, the velocity dispersion is fitted separately in 2D following a basic power law,
\begin{equation}
    \sigma^2_{\rm{los}}(r) = \alpha r^{\beta}.
\end{equation}
This is used alongside the parameters found from the SL temperature model to obtain a fit for the $\alpha$ profile over the limited range (0.5-1.5)$R_{500}$, as we only require it to estimate the mass correction at $R_{500}$.

Fig. \ref{fig:halos_103_51_51} shows emission measure (left column), SL temperature (middle) and velocity dispersion (right) maps of three clusters. The top row shows a perturbed (see Section \ref{sec:morph} for classification criteria), relatively low mass ($M_{500} = 7.0 \times 10^{14} \mathrm{M}_{\odot}$) cluster at $z=0$. The middle and bottom rows show a second cluster at $z=0.5$ and $z=0$ respectively. At $z=0.5$ the cluster is perturbed with a mass of $7.4 \times 10^{14}$ $\mathrm{M}_{\odot}$ and at $z=0$ it is relaxed with a higher mass of $1.4 \times 10^{15}$  $\mathrm{M}_{\odot}$. Each map has a total side length of 6 $R_{500}$, they have been smoothed using \texttt{swiftsimio} \citep{Borrow2020} and have a pixel size of 0.01 $R_{500}$. The maps are normalised to their average value, the lighter colours represent higher emission, hotter temperature or higher dispersion. The disturbed clusters have more substructures and are more irregularly shaped in the emission measure maps, especially within $R_{500}$ (the radius used to determine the dynamical state). However, there is little qualitative difference between the temperature and velocity dispersion maps for relaxed and perturbed clusters, other than a few more cold spots in the latter case.

\subsubsection{Clumping and azimuthal scatter}

To measure the clumping from a particle data set resulting from simulations with smoothed particle hydrodynamics, we use,
\begin{equation}
    \mathscr{C} = \frac{ \sum_{i} m_{i} \rho_{i} \sum_i m_{i} / \rho_{i}} {(\sum_{i} m_{i})^{2}}
    \label{eq:clumping_particles}
\end{equation}
\citep{Battaglia2015ONSUBSTRUCTURES, Planelles2017}. The clumping was measured for each cluster using spherical shells. The same temperature cut as the projected profiles was used to exclude the coolest and densest gas clumps as we only want to investigate fluctuations in the hot ionised gas.

Thus it is a purely theoretical quantity which observers are unable to measure. Instead, observational proxies have been suggested to estimate the level of clumping in a cluster. In this work we have used the azimuthal scatter, defined as
\begin{equation}
    \sigma_{A}(r) = \sqrt{\frac{1}{N} \sum_{i=1}^{N} \left( \frac{X_{i}(r)-\langle X(r)\rangle}{\langle X(r)\rangle} \right) ^{2}}
    \label{eq:az_scatter}
\end{equation}
following \citet{Vazza2011}, in both the emission measure and SL temperature to measure the level of gas inhomogeneities within a cluster. We split each radial bin into $N=12$ angular bins and measure the emission measure/SL temperature in each angular bin, $X_{i}(r)$, and compare this to the overall median at the relevant radius, $\langle X(r) \rangle$.

\begin{figure}
    \centering
    \includegraphics[scale=0.55]{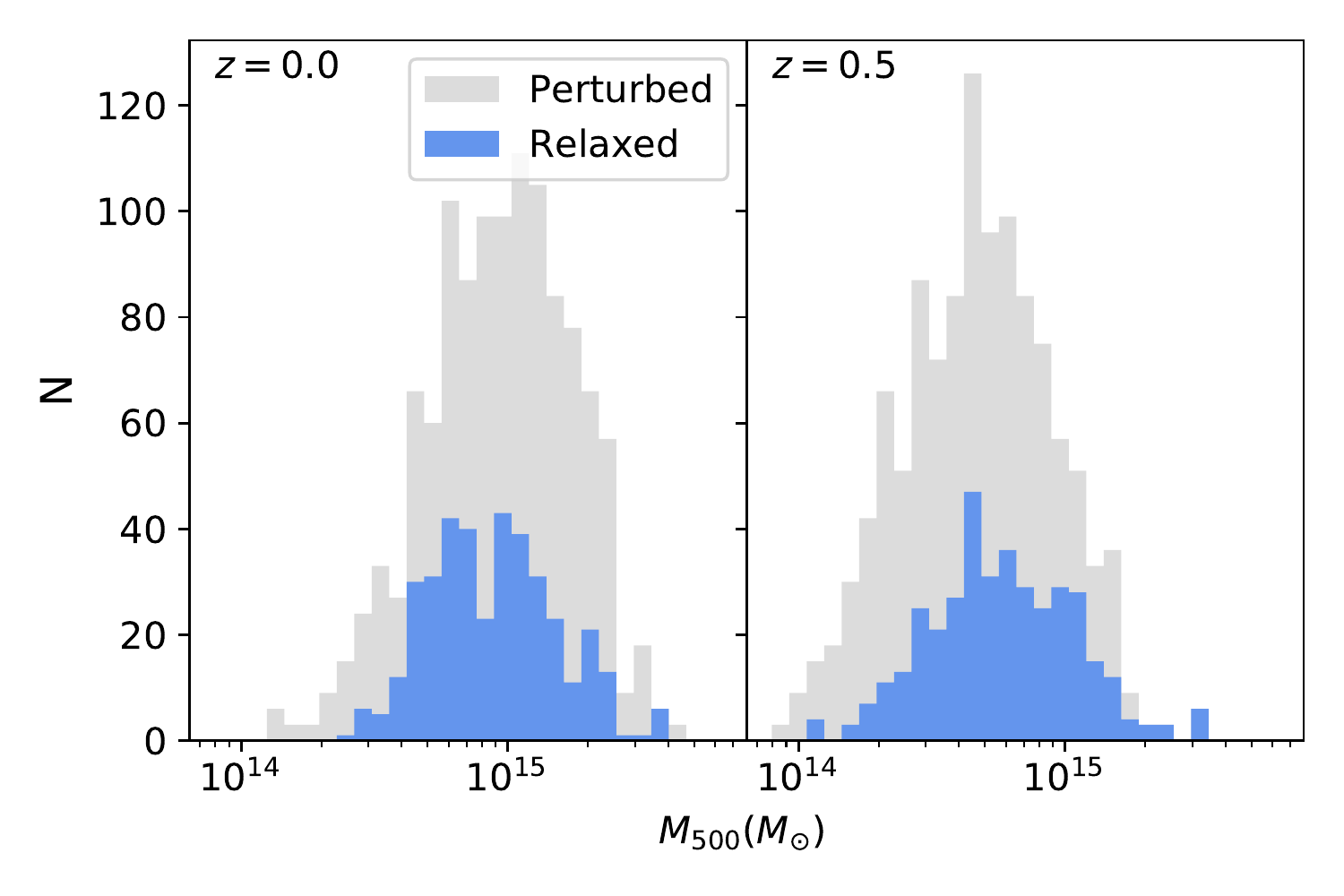}
    \caption{Distribution of cluster mass at $z=0$ (left) and $z=0.5$ (right). The relaxed clusters are shown in blue and the perturbed in grey. The median mass is $\langle M_{500} \rangle = $ 9.5$\times 10^{14} \rm{M}_{\odot}$ at $z=0$ and 4.7 $\times 10^{14} \rm{M}_{\odot}$ at $z=0.5$. The dynamical classification of the cluster was made for each of the three perpendicular projections and so all three are shown on this plot to accurately describe the distribution of regular and perturbed clusters. Hence, we show 1131 clusters in total.}
    \label{fig:mass_dist_class}
    %code: mass_histograms.py
\end{figure}

\subsection{Dynamical state classification}
\label{sec:morph}

The dynamical states of the clusters were classified using emission measure maps (as a proxy for X-ray surface brightness). Properties such as the symmetry, overall shape and distribution of brightness in the cluster were analysed to determine whether a cluster was perturbed or relaxed. This gave four morphology statistics in total: the surface brightness concentration, symmetry statistic, alignment statistic and centroid shift.

\begin{figure*}
    \centering
    \includegraphics[trim={0 2.5cm 0 1cm}, clip]{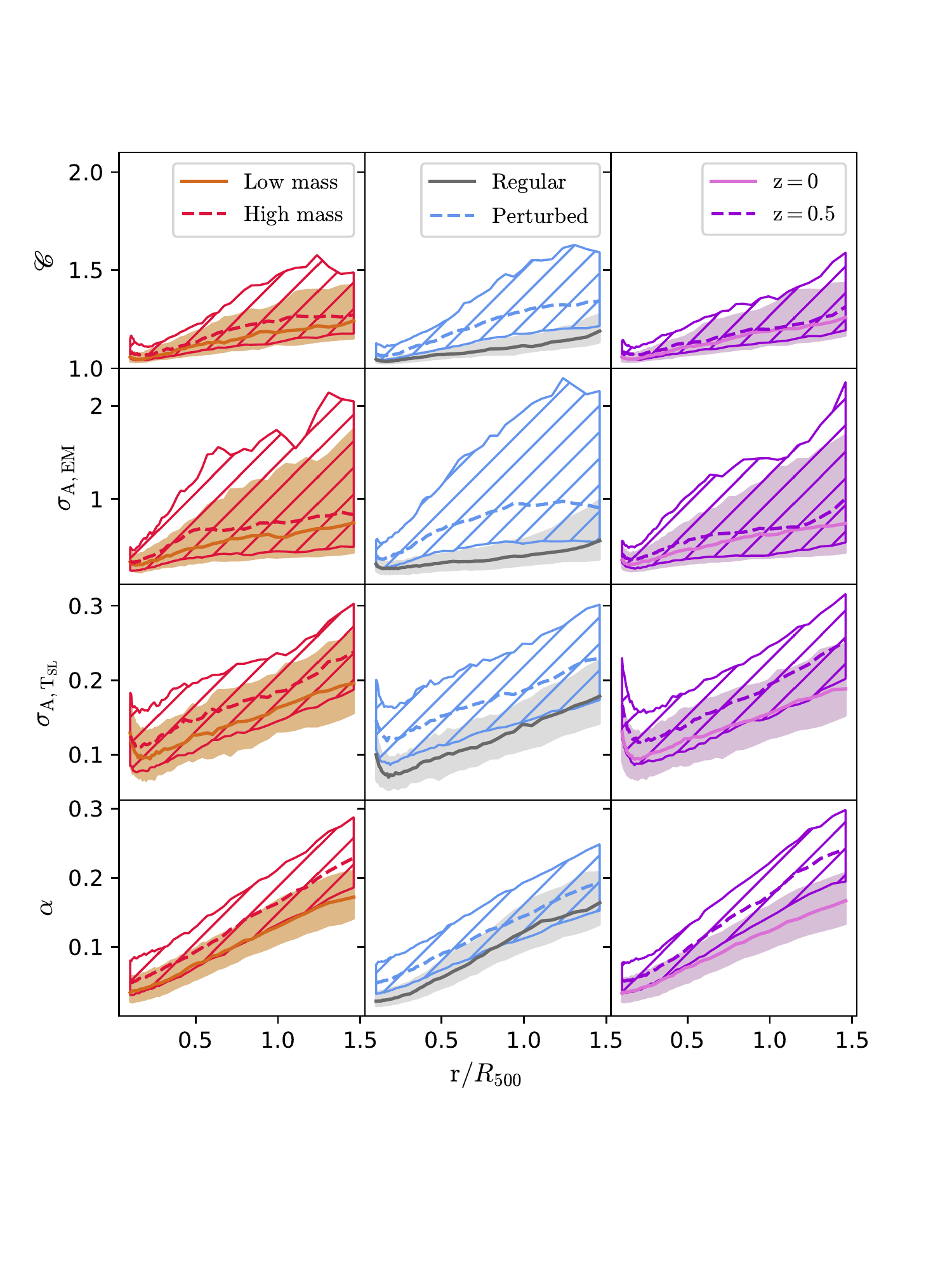}
    \caption{Median profiles for the clumping (top), azimuthal scatter in the emission measure (second), azimuthal scatter in the SL temperature (third) and projected non-thermal pressure (bottom) in the MACSIS clusters. The left panel includes MACSIS clusters with $M_{500} > 10^{14.78} \rm{M}_{\odot}$ at redshift $z=0$. The central panel uses data from clusters in the MACSIS analysis sample at $z=0$. The rightmost plot compares clusters with $10^{14.55} M_{\odot} < M_{500} < 10^{15.1} M_{\odot}$ at redshifts $z=0$ and $z=0.5$. The shaded areas denote upper and lower quartiles in the sample.}
    \label{fig:clump_obs_profs}
    %code: HSE_property_profiles.py
\end{figure*}

The surface brightness concentration, $c$, identifies clusters with a brighter, cooler core \citep[which tend to be more relaxed,][]{Peterson2006} by measuring the fraction of X-ray emission ($SB$) that comes from the core of a cluster via
\begin{equation}
    c = \frac{SB(<0.15 R_{500})}{SB(<R_{500})}.
\end{equation}
A cluster is classified as relaxed if $c >0.5$. 

For the symmetry and alignment statistics, we follow the procedure laid out by \citet{Mantz2015}, where a series of $N_{\rm{el}}=5$ ellipses are fitted to isoflux contours at evenly spaced intervals within 0.15 - 1.0 $R_{500}$ and their centres calculated. The symmetry statistic, $s$, measures how much these fitted centres shift relative to the global centre of the cluster via
\begin{equation}
    s = - \log_{10} \left( \frac{1}{N_{\rm{el}}} \sum_{j=1}^{N_{\rm{el}}} \frac{\delta_{j,c}}{\langle b_{\rm{el}} \rangle_j} \right),
\end{equation}
where $\delta_{j,c}$ represents the distance between the centre of the $j$th fitted ellipse and the global centre and $\langle b_{\rm{el}} \rangle_j$ is the average of the major and minor axes of the same ellipse. This gives an estimate for the asymmetry of the cluster. 
If $s>0.87$, the cluster is relaxed. 

Similarly, the alignment statistic looks at how the centres shift relative to the adjacent ellipses, therefore measuring the level of substructure in a cluster.It is defined as
\begin{equation}
    a = - \log_{10} \left( \frac{1}{N_{\rm{el}} - 1} \sum_{j=1}^{N_{\rm{el}}-1} \frac{\delta_{j,j+1}}{\langle b_{\rm{el}} \rangle_{j,j+1}} \right),
\end{equation}
where $\delta_{j,j+1}$ measures the distance between centres of adjacent fitted ellipses and $\langle b_{\rm{el}} \rangle_{j,j+1}$ is the average of the major and minor axes of both ellipses. A cluster is classified as relaxed if $a>1$.

Finally, the centroid shift, $\langle w \rangle$, is calculated following \citet{Maughan2012}, 
\begin{equation}
    \langle w \rangle = \frac{1}{R_{500}} \sqrt{\frac{\sum\left( \Delta_{i} - \langle \Delta \rangle \right)^2}{M-1}}.
\end{equation}
We measure how the centroid of a cluster, $\Delta_{i}$, within a series of $M=8$ increasing smaller apertures within $0.15 - 1.0 R_{500}$, changes relative to the average $\langle \Delta \rangle$. The centroid shift quantifies how regularly shaped a cluster is; relaxed clusters are identified if $\langle w \rangle < 0.006$.

Three maps for each cluster were used, each one projected down a line of sight perpendicular to the others, i.e. \{x, y, z\}, giving each cluster one morphology classification for each projection. A cluster projection would be classified as relaxed if three or more of the morphology statistics measured it to be relaxed, otherwise it was classified as perturbed. Fig. \ref{fig:mass_dist_class} shows the distribution of the masses of clusters in the MACSIS sample, in addition to the distribution of relaxed (blue) and perturbed clusters (grey). Around two thirds of the clusters were defined to be perturbed at $z=0$, increasing to around 80 per cent at $z=0.5$. We also find that at $z=0$ perturbed clusters are, on average, more massive, but the opposite is found at $z=0.5$, however the effect is smaller at higher redshifts.

\begin{figure*}
    \centering
    \includegraphics{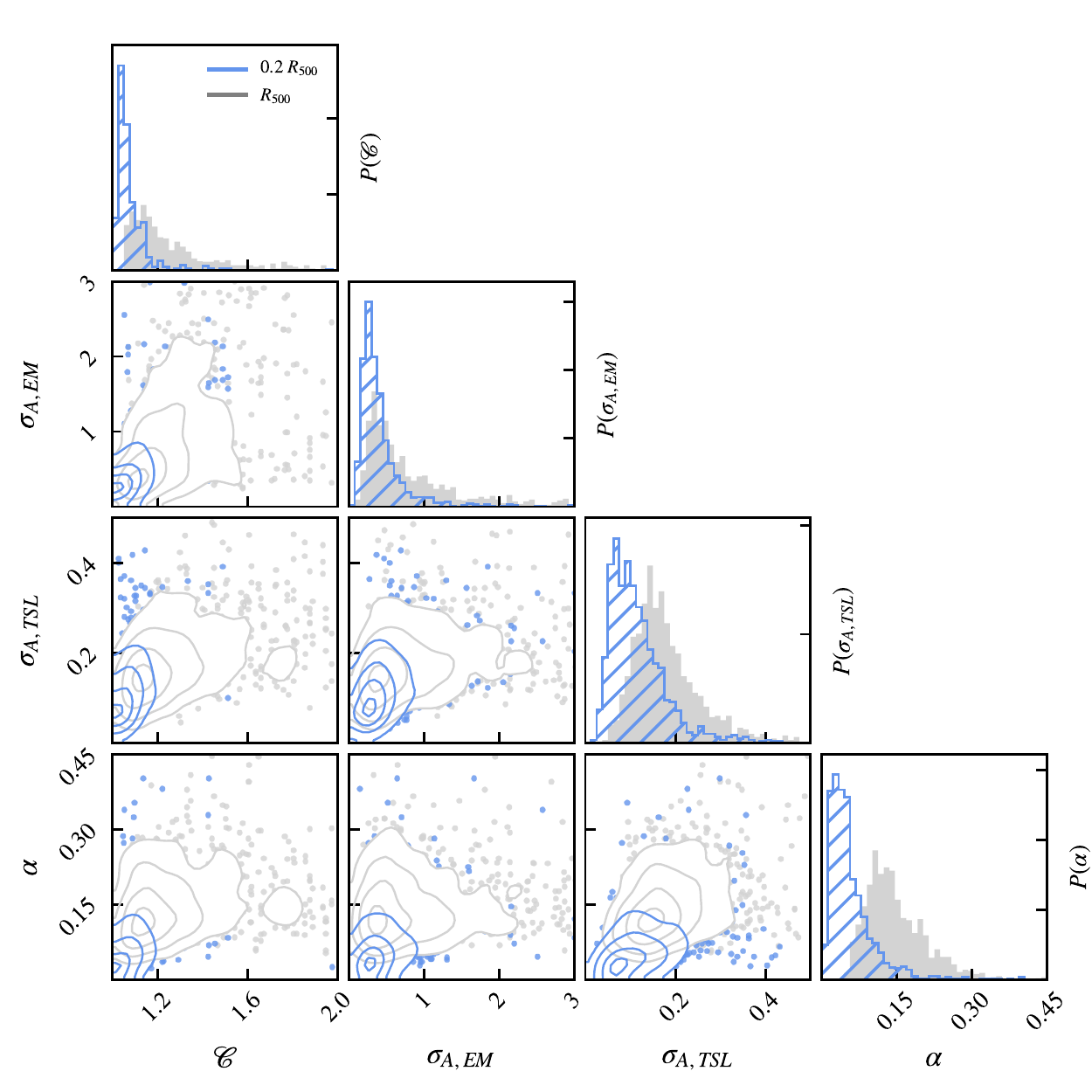}
    \caption{Correlations between the spherical clumping and the azimuthal scatter of the emission measure and spectroscopic-like temperature, and the non-thermal pressure fraction at radii of 0.2 $R_{500}$ (blue) and 1.0 $R_{500}$ (grey). The distributions of these values are shown in the diagonal panels.} 
    %code: made by Edo
    \label{fig:clump_props_corner}
\end{figure*}

\section{Inhomogeneities in the ICM and their effect on radial profiles}
\label{sec:inhomogeneities}

In this section, we assess the amount of ICM clumping in the MACSIS clusters, its correlation with potential X-ray observable proxies, and how well using the azimuthal median rather than the mean reduces the effects of gas clumping in projected radial profiles.
  
\subsection{Clumping and observational proxies}

The median 3D clumping profiles for the MACSIS clusters are shown in the top panels of Fig. \ref{fig:clump_obs_profs}, where the left column splits the sample into low (orange) and high (red) mass clusters with a mass cut, $M_{500} < 10^{14.78}$ $\rm{M}_{\odot}$, to remove the lowest mass objects which tend to be irregular and underconcentrated \citep{Henson2017}. In total 82/ 377 clusters are removed, with the low and high mass bins divided by the median mass, $M_{500}=10^{15.07}M_{\odot}$. The middle column splits the sample of 377 into relaxed (grey) and perturbed (blue) clusters and the right column shows the clumping profiles for clusters at redshifts $z=0$ (solid, 232 clusters in total) and $z=0.5$ (dashed, 230 clusters in total) within the same mass range of $10^{14.55} \rm{M}_{\odot} < M_{500} < 10^{15.1} \rm{M}_{\odot}$.

The clumping in these clusters agree with other works that the clumping increases in the outskirts of clusters \citep{Nagai2011GASCLUSTERS, Vazza2013PropertiesMedium, Zhuravleva2013, Roncarelli2013, Khedekar2013BiasObservations, Eckert2015, Battaglia2015ONSUBSTRUCTURES, Planelles2017, Ansarifard2020}. We also find that more massive and disturbed clusters tend to have more gas clumps and the clumping within a fixed mass range is approximately independent of redshift (to $z=0.5$).
As expected, perturbed clusters have more gas inhomogeneities than relaxed clusters as well as a larger spread in clumping values. Perturbed clusters are more likely to have substructures and be more elliptical, both of which increase the clumping. For example, the cluster in the middle row of Fig. \ref{fig:halos_103_51_51} is perturbed and has qualitatively more substructure, giving it a higher clumping than its relaxed descendent shown in the bottom row.

Clumping is not a directly measurable quantity, instead observational proxies such as the azimuthal scatter \citep{Vazza2011, Eckert2012, Roncarelli2013, Ansarifard2020} have been proposed to quantify the magnitude of gas inhomogeneities. The median 2D profiles of the azimuthal scatter in the emission measure are shown in the second row of Fig. \ref{fig:clump_obs_profs}. The trends closely match that of the clumping: the scatter increases with radius, is much higher in perturbed clusters and is relatively unaffected by redshift. In addition, the third row showing the 2D azimuthal scatter in the temperature and the bottom row showing the 2D non-thermal pressure fraction also increase with radius. In contrast to the clumping, these two properties also have a redshift dependence, with higher redshift clusters giving a higher temperature scatter and non-thermal pressure fractions. Note the relatively low amount of non-thermal pressure at $R_{500}$ ($\alpha \approx 0.1-0.15$) due to the subtraction of local bulk motion in each azimuthal bin (see also \citealt{Angelinelli2020TurbulentMedium} and \citealt{Bennett2021ABiases}).  

\begin{table}
    \centering
    \caption{Spearman correlation coefficients and respective bootstrap errors between the clumping and various observables (non-thermal pressure fraction and azimuthal scatter in the emission measure and temperature) at two fixed radii.}
    \begin{tabular}{|l|l|l|}
    \hline
        $r_{s}$ & $0.2 R_{500}$ & $1.0 R_{500}$\\ 
        \hline
        $\mathscr{C}, EM$ ~        & 0.67 $\pm$ 0.02 & 0.65 $\pm$ 0.02 \\ %0.39 $\pm$ 0.08 & 0.2  $\pm$ 0.1  \\
        $\mathscr{C}, T_{\rm{SL}}$ & 0.57 $\pm$ 0.02 & 0.51 $\pm$ 0.02 \\ %0.48 $\pm$ 0.06 & 0.11 $\pm$ 0.04 \\
        $\mathscr{C}, \alpha$      & 0.63 $\pm$ 0.02 & 0.32 $\pm$ 0.03 \\ %0.35 $\pm$ 0.04 & 0.11 $\pm$ 0.03 \\
        $\alpha, EM$               & 0.44 $\pm$ 0.03 & 0.14 $\pm$ 0.03 \\ %0.20 $\pm$ 0.03 & 0.12 $\pm$ 0.03 \\
        $\alpha, T_{\rm{SL}}$      & 0.54 $\pm$ 0.02 & 0.36 $\pm$ 0.03 \\ %0.37 $\pm$ 0.03 & 0.39 $\pm$ 0.03 \\
        $T_{\rm{SL}}, EM$          & 0.51 $\pm$ 0.02 & 0.46 $\pm$ 0.02 \\ %0.43 $\pm$ 0.09 & 0.17 $\pm$ 0.04 \\
        \hline
    \end{tabular}
    \label{table:obs_corr}
\end{table}

The clumping, azimuthal scatter in the emission measure and temperature, and non-thermal pressure fraction of individual clusters are compared directly in Fig. \ref{fig:clump_props_corner}, at both 0.2 $R_{500}$ (blue) and $R_{500}$ (grey),the quantities are measured in radial bins of range $0.18-0.22 R_{500}$ and $0.95-1.05 R_{500}$. Each diagonal panel shows the probability distribution of the value, while the off diagonal panels show the correlation between quantities on the x and y-axis. The contours show the general correlation with the points showing outlying clusters. 

We find a Spearman correlation coefficient of $r_{s} = 0.67 \pm 0.02$ between the clumping and the scatter in the emission measure at 0.2 $R_{500}$ and $r_{s} = 0.65 \pm 0.02$ at $R_{500}$, showing a correlation between the two quantities. The full list of correlation coefficients is shown in Table \ref{table:obs_corr}. Correlations are significant between all quantities, but are weaker at the larger radius.

\citet{Roncarelli2013} found a similar correlation between the clumping and azimuthal scatter in the surface brightness, with a Spearman correlation coefficient of $r_{s} = 0.6$ for a sample only including the relaxed clusters, and used this strong correlation to use the azimuthal scatter to estimate the clumping. However, their relation uses residual clumping, which requires an elimination of the one per cent densest gas. In this work, the need for using the residual clumping was removed by using a temperature cut, which removes the coldest, densest clumps of gas and eliminates the possibility of getting rid of the largest density fluctuations in the hot gas that may be strongly influencing the density profiles. \citet{Ansarifard2020} show similar results with a correlation of $r_{s} = 0.56$ between standard (not residual) clumping and azimuthal scatter in the surface brightness at $R_{500}$.

\subsection{Impact of clumping on radial gas profiles}
\label{sec:radial}

\begin{figure}
    \centering
    \includegraphics[trim=0 0 0 0, clip]{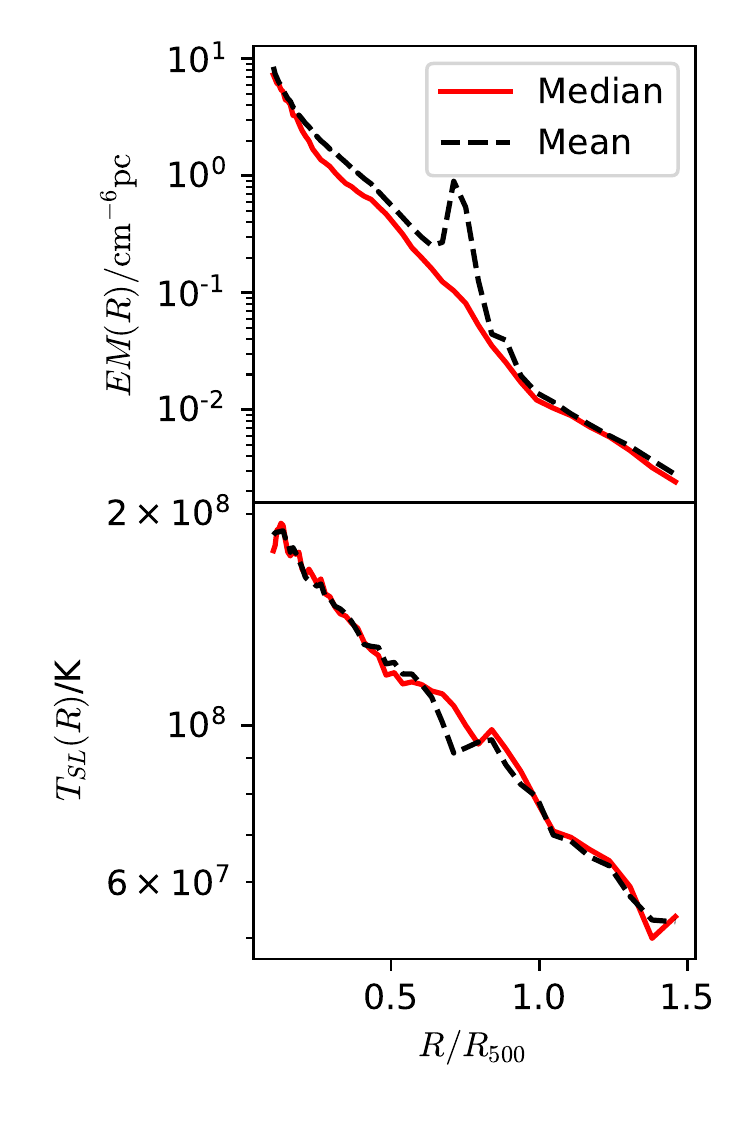}
    \caption{2D radial profiles for an example MACSIS cluster. The profiles are calculated using 12 angular sectors at each radius. The mean (median) profile is calculated using the 12 bins and is shown as the black, dashed (red, solid) line.}
    \label{fig:mean_median_20}
    %code: median_mean_property_profiles.py
\end{figure}

\begin{figure}
    \centering
    \includegraphics[scale=0.65, trim={0.6cm 0 1.5cm 0.8cm}, clip]{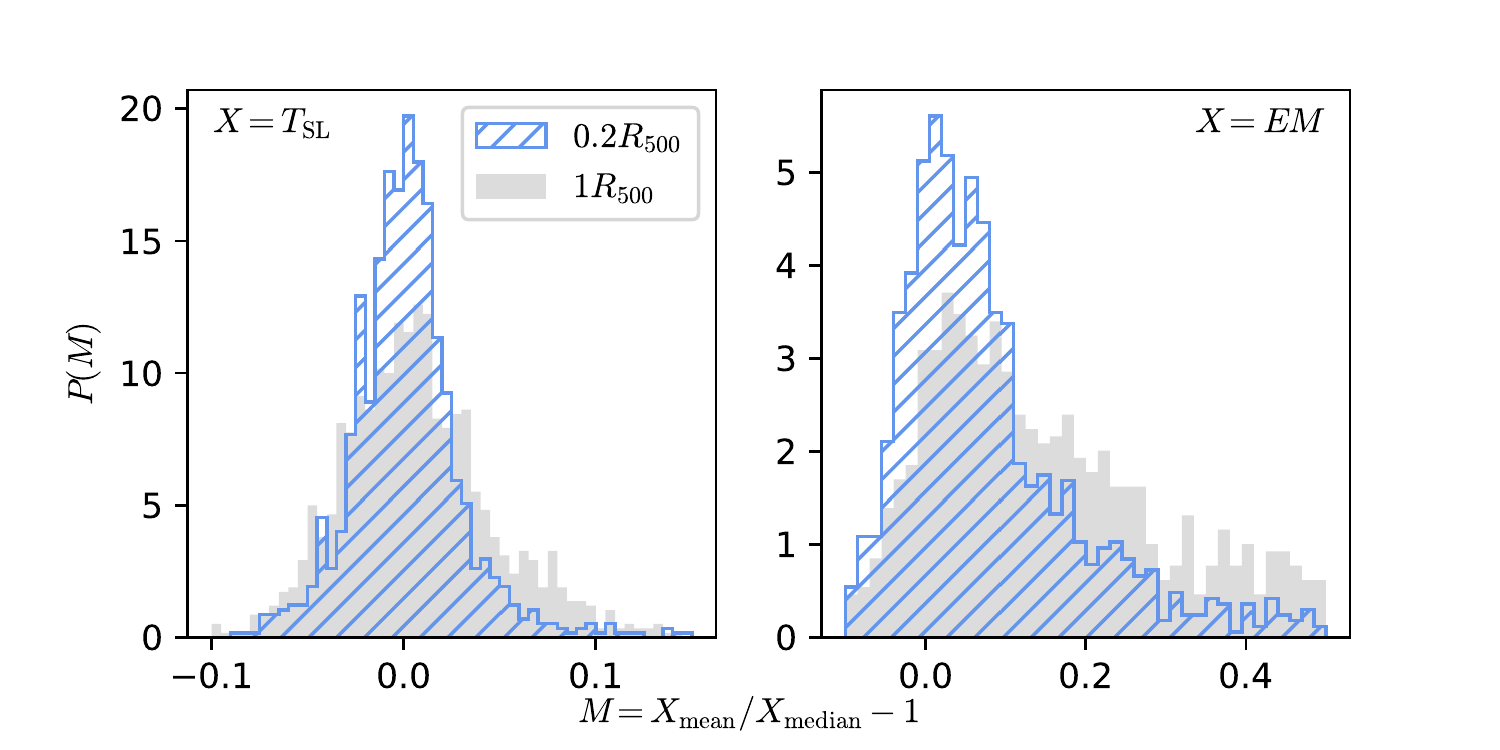}
    \caption{Distribution of mean/median values determined at both 0.2 (blue) and 1 (grey) $R_{500}$ of emission measure (right) and SL temperature (left).}
    \label{fig:MM_hist}
    %code: clumping_v_properties_fixed_r.py
\end{figure}

Current X-ray observations cannot resolve small-scale ICM fluctuations but future surveys are expected to improve on this \citep{Morandi2013Non-parametricClusters, Eckert2015}. Therefore, instead of measuring the gas inhomogeneities directly, some works have removed the effect of clumping from their profiles by eliminating denser clumps from simulations or brighter spots from observations \citep[e.g.][]{Vazza2011,Vazza2013PropertiesMedium, Khedekar2013BiasObservations, Roncarelli2013}. However, \citet{Zhuravleva2013} found that mean gas property profiles were sensitive to the methods used to remove the clumps. They proposed the use of azimuthal median profiles rather than the mean \citep[the median is also used in][]{Eckert2015}, which was found to be more robust in the presence of gas inhomogeneities. 

To illustrate the effects of clumping on MACSIS radial profiles, Fig. \ref{fig:mean_median_20} shows the mean (black, dashed) and median (red, solid) profiles of the emission measure (top) and spectroscopic-like temperature (bottom) for an example cluster. The cluster contains a bright substructure, causing a peak in the mean, but not median, emission measure profile at 0.7-0.8$R_{500}$.

\begin{table*}
    \centering
    \caption{The median hydrostatic mass bias values with their respective standard deviation, $\sigma_{b_{\rm{HSE}}},$ for measurements using the 3D unprojected profiles, all the 2D projected profiles, and the relaxed and perturbed subsamples of the projected profiles. The table also includes the median bias values and scatter after the non-thermal pressure correction term is included.}
    \begin{tabular}{|c|c|c|c|c|}
    \hline
         ~ & \multicolumn{2}{|c|}{Uncorrected} & \multicolumn{2}{|c|}{$\alpha$ Corrected}  \\ 
         ~ & $1-b_{\mathrm{HSE}}$ & $\sigma_{b_{\mathrm{HSE}}}$ & $1-b_{\mathrm{HSE}}$ & $\sigma_{b_{\mathrm{HSE}}}$ \\ \hline
         True 3D profiles                 & 0.86  $\pm$ 0.01  & 0.24  $\pm$ 0.01 & 1.16 $\pm$ 0.03 & 0.39 $\pm$ 0.02\\
         Projected 2D profiles, all        & 0.753 $\pm$ 0.006 & 0.171 $\pm$ 0.005 & 0.851 $\pm$ 0.006 & 0.262 $\pm$ 0.009\\ 
         Projected, Relaxed               & 0.802 $\pm$ 0.007 & 0.120 $\pm$ 0.006 & 0.854 $\pm$ 0.008 & 0.16 $\pm$ 0.01 \\ 
         Projected, Perturbed             & 0.721 $\pm$ 0.006 & 0.187 $\pm$ 0.007 & 0.85 $\pm$ 0.01 & 0.31 $\pm$ 0.01 \\ \hline
    \end{tabular}
    %code: HSE_projected_deprojected.py
    \label{table:bias_vals}
\end{table*}

The distribution of the mean-median ratio, defined as $M=X_{\rm{mean}}/X_{\rm{median}}-1$, for SL temperature (left) and emission measure (right) is shown in Fig. \ref{fig:MM_hist}. The gas inhomogeneities (such as the peak in Fig. \ref{fig:mean_median_20}) cause the mean emission measure profile to be larger than the median, giving a long tail in its distribution, with medians of 0.05 and 0.2 at 0.2 $R_{500}$ and $R_{500}$ respectively. The SL temperature profile in Fig. \ref{fig:mean_median_20} shows little difference between the mean and median. This is reflected in Fig. \ref{fig:MM_hist} where $M$ is centred around zero with much smaller scatter (with medians of 0.001 and 0.005 for 0.2 $R_{500}$ and $R_{500}$ respectively), i.e. the SL temperature is not particularly sensitive to gas inhomogeneities on the scales we have used to define our azimuthal bins.  Note that, in both cases, the $M$ distribution is wider at larger radii, reflecting the larger fluctuations there (as shown in Fig. \ref{fig:clump_obs_profs}).

\section{Clumping and hydrostatic mass estimates}
\label{sec:mass_bias}

Measuring the hydrostatic mass of a cluster requires the assumption that a cluster is in hydrostatic equilibrium. This introduces a mass bias, which we aim to reduce,or correct for, in this section using the level of gas clumping and non-thermal pressure within a cluster.

We use models from \citet{Vikhlinin2006} to fit the gas density and temperature profiles as mentioned in section \ref{sec:profs}. These are then used to calculate the hydrostatic mass via
\begin{equation}
    M_{\mathrm{HSE}}(<r) = -\frac{k_{B} T(r) r}{\mu m_{\rm{H}}} \left( \frac{\mathrm{d}\log T}{\mathrm{d}\log r} + \frac{\mathrm{d}\log \rho}{\mathrm{d}\log r}\right).
    \label{eq:HSE}
\end{equation}
We then interpolate to find the radius at which the average density is 500 times the critical density, $R_{500,\rm{HSE}}$, and hence calculate the final mass, $M_{\rm{HSE},500}$ and bias, $b = 1 - M_{500,\rm{HSE}}/M_{500, \rm{true}}$. Other works discussing the hydrostatic mass bias using simulations often compare the hydrostatic mass bias at $R_{500}$ determined from the simulation to compare bias at a fixed radius \citep[for example][]{Ansarifard2020, Gianfagna2021ExploringSample}. However, as the true scale is not known in observations, we have computed the mass at $R_{500,\rm{HSE}}$ to include this additional uncertainty \citep[similarly][also determined the mass bias at this radius]{Henson2017, Barnes2017}.

\begin{figure}
    \centering
    \includegraphics[scale=0.62]{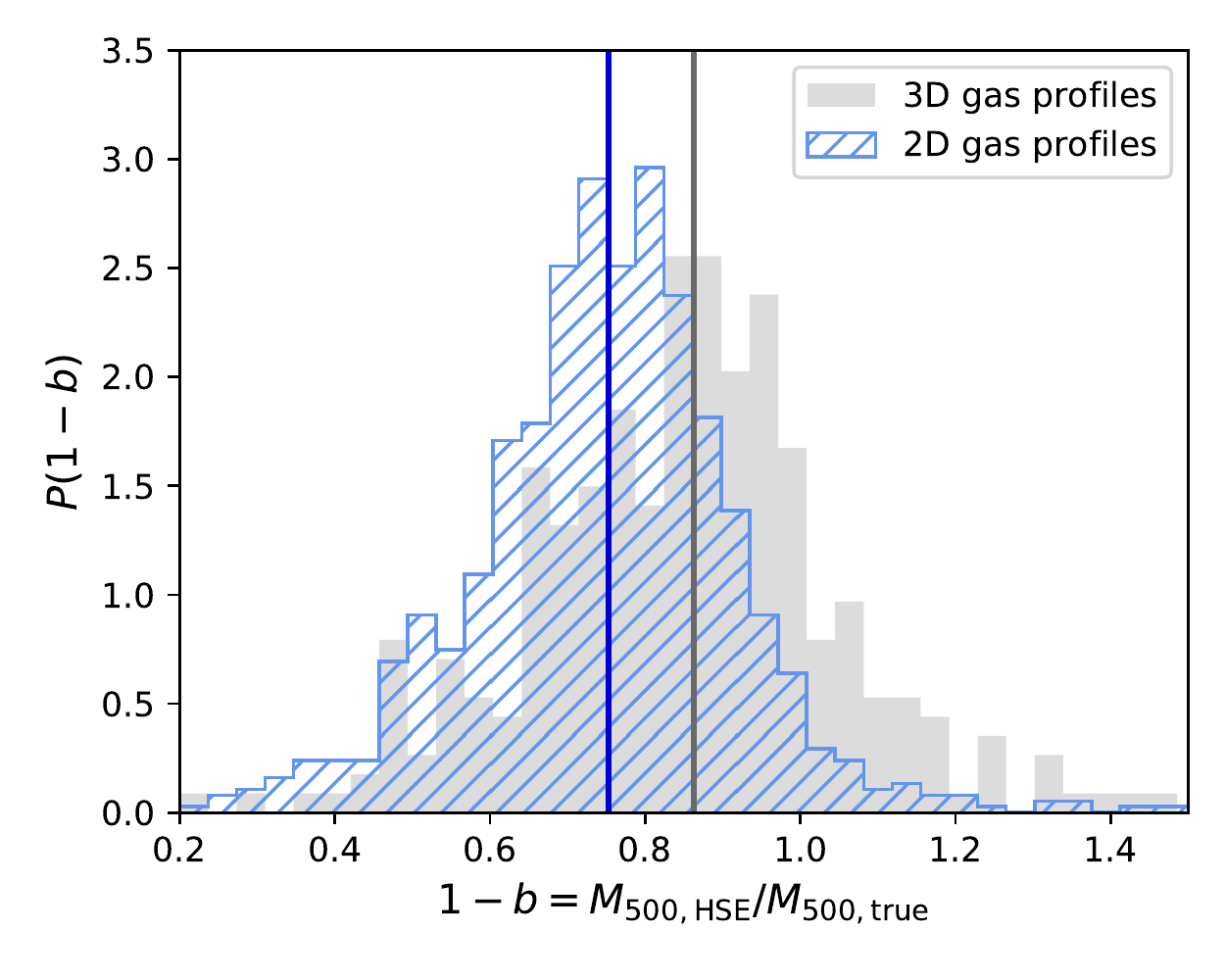}
    \caption{The distribution of HSE mass bias values calculated using both 3D (solid, grey) and 2D gas profiles (hatched, blue). The vertical lines denote the median of each distribution.}
    \label{fig:HSE_mass_2D_3D}
    %code: HSE_projected_deprojected.py
\end{figure}

The distribution of mass bias calculated in this work is shown in Fig. \ref{fig:HSE_mass_2D_3D}, using both 3D profiles (solid, grey) and projected 2D profiles (dashed, blue). The 3D bias we obtained agrees with what has been found previously, with a median $1-b=0.86 \pm 0.01$. We find that for the projected profiles the bias is lower, $1-b=0.753 \pm 0.006$, but the scatter remains similar (see Table \ref{table:bias_vals}) to the 3D result.
This matches what \citet{Henson2017}, \citet{Pearce2020} and \citet{Barnes2021CharacterizingMock-X} found for the MACSIS (and similar CELR-B) simulations, i.e. that the bias is significantly reduced by using true gas density and temperature profiles when compared to spectroscopic profiles. \citet{Henson2017} identified the main cause of this to be due to the bias in the spectroscopic temperature they have used. A similar bias exists in the spectroscopic-like temperature profile used in this work, where the median spectroscopic-like temperature of all clusters is approximately 22 per cent
smaller than the median mass weighted temperature at $R_{500,\rm{true}}$ and so, following Equation \ref{eq:HSE}, will bias the HSE mass low.

To compare whether using the azimuthal median to extract gas property profiles affects the hydrostatic mass bias, hydrostatic mass estimates were also made using mean gas property profiles. We find that the mean reduces the bias slightly in comparison to the azimuthal median ($1-b = $ 0.771 $\pm$ 0.007 for the mean and 0.753 $\pm$ 0.006 for the median), however the scatter in the distribution of mass bias values is larger (0.23 $\pm$ 0.01 for the mean and 0.171 $\pm$ 0.005 for the median). The average bias values have also been calculated for the relaxed and perturbed cluster subsets and we find that the relaxed clusters have a lower bias on average with a smaller scatter.

\begin{figure}
    \centering
    \includegraphics[scale=0.7, trim=0 0 1cm 0, clip]{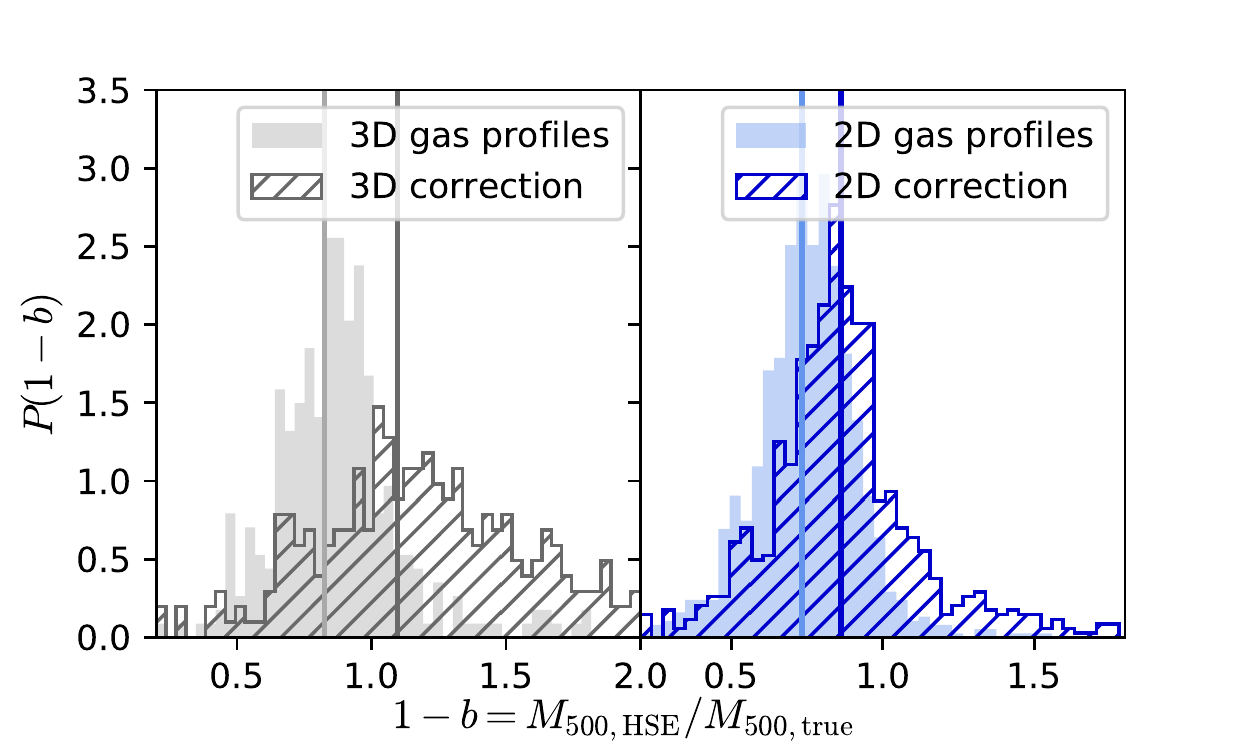}
    \caption{The distribution of HSE mass bias values. Left (right) is calculated using 3D (2D) gas profiles. The solid histogram refers to the uncorrected HSE mass estimates, while the dashed shows the $\alpha$-corrected masses.}
    \label{fig:HSE_mass_corr_dists}
    %code: HSE_projected_deprojected.py
\end{figure}

\begin{figure*}
    \centering
    \includegraphics[trim=0.45cm 0 1cm 0.6cm, clip, scale=0.9] {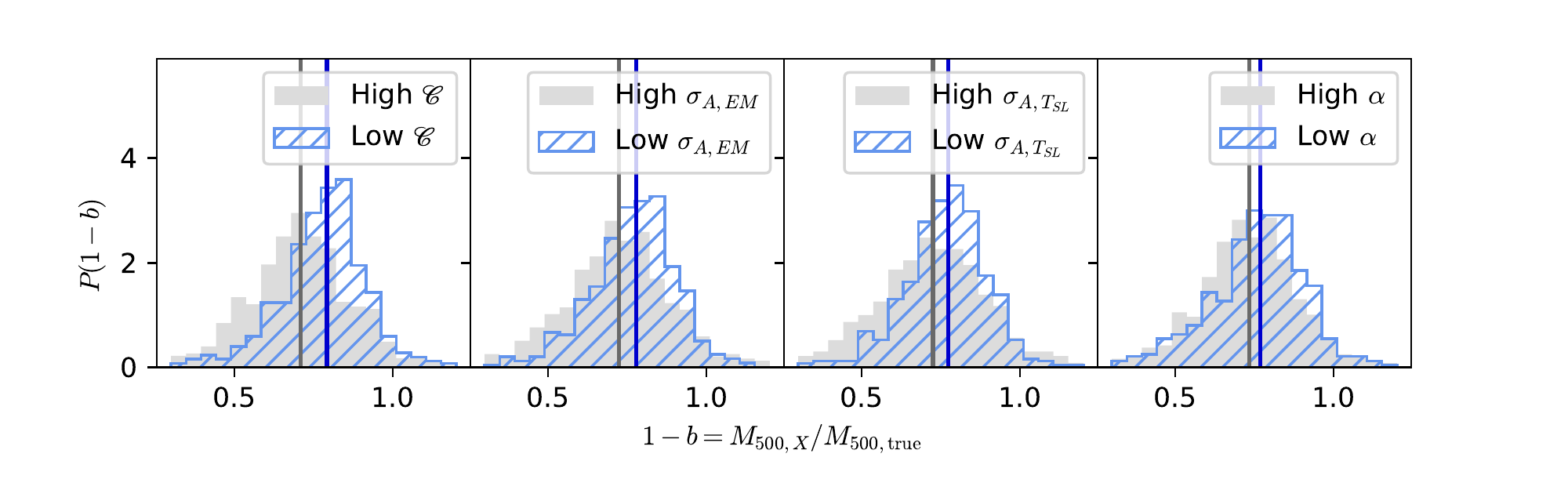}
    \caption{The bias probability distribution function when grouping the cluster sample into different bins. On the top left, clusters are split into high and low values for the clumping measured at R$_{\rm{HSE,500}}$. They are also split according to the azimuthal scatter in the emission measure (top right), temperature (bottom left) and velocity dispersion (bottom right) at the same radius}. The sample excludes clusters for which M$_{\rm{HSE},500}$ could not be calculated and therefore R$_{\rm{HSE},500}$ could not be estimated. The clusters are split such that each high and low sample is approximately equal (i.e. $\mathscr{C}=1.2, \sigma_{\rm{A,EM}}=0.63, \sigma_{\rm{A,T_{SL}}}=0.155, \alpha=0.13$). The vertical lines show the median of the distributions.
    \label{fig:bias_pdfs}
    %code: HSE_bias_obs.py
\end{figure*}

\begin{figure*}
    \centering
    \includegraphics[trim=0.45cm 0 1cm 0.6cm, clip, scale=0.9] {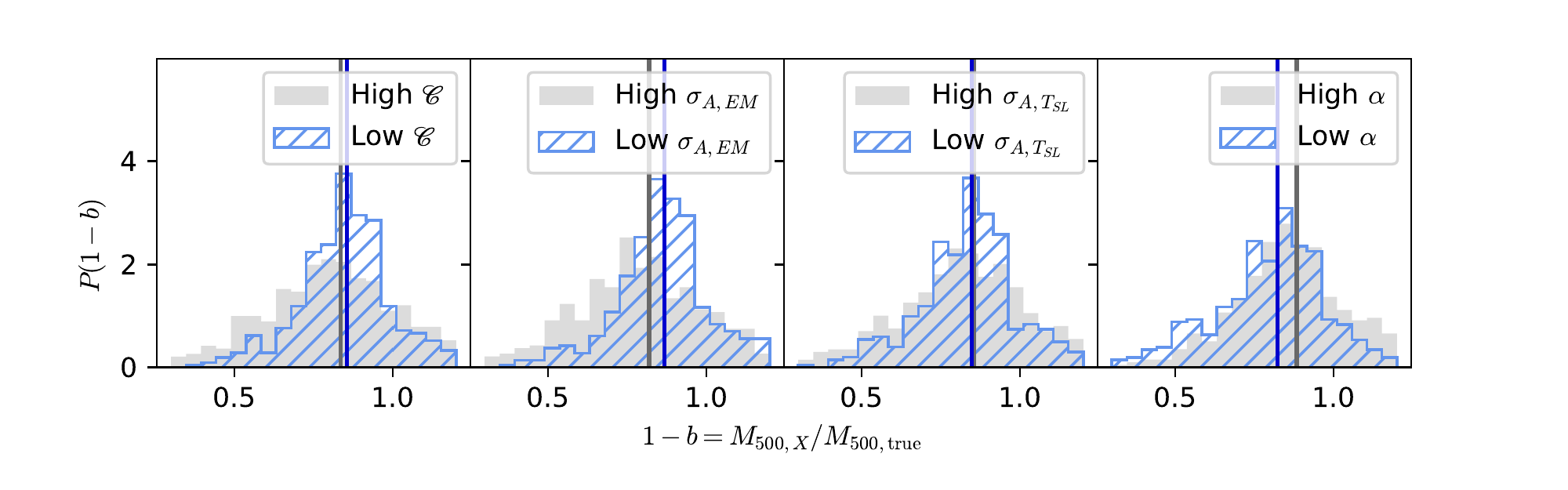}
    \caption{The non-thermal pressure corrected bias probability distribution function when splitting the cluster sample into different groups. The criteria for sampling are the same as in Fig. \ref{fig:bias_pdfs}. The vertical lines show the median of the distributions.}
    \label{fig:corr_bias_pdfs}
    %code: HSE_bias_obs.py
\end{figure*}

\subsection{Non-thermal pressure corrections to the mass estimate}
\label{sec:non-therm}

To reduce the bias caused by gas motions, previous works have incorporated the effect of non-thermal pressure into their mass estimates. Both  \citet{Shi2016} and \citet{Pearce2020} reduced the mass bias by approximately 20 per cent with this method, and \citet{Lau2009ResidualClusters} found that the cluster mass accuracy can also be improved by accounting for the non-thermal pressure.
The non-thermal pressure is used to give a corrected mass estimate, following
\begin{equation}
    M_{\mathrm{HSE,corr}}\left(<r\right) = \frac{1}{1 - \alpha} 
    \left[ M_{\mathrm{HSE}} - \frac{\alpha}{1 - \alpha} \frac{k_{B}Tr}{G\mu m_{\rm{H}}} \frac{\mathrm{d} \log \alpha}{\mathrm{d} \log r}\right],
    \label{eq:HSE_corr}
\end{equation}
where $\alpha$ is defined in Equation \ref{eq:alpha}.

Fig. \ref{fig:HSE_mass_corr_dists} shows the distribution of $\alpha$-corrected mass estimates. The left panel was obtained using unprojected 3D gas profiles, while the right used projected 2D profiles. Both solid histograms show uncorrected distributions and hatched shows the corrected values. The correction is stronger in 3D profiles (see Table \ref{table:bias_vals}), giving a negative bias, with a wider scatter, driven by the more perturbed clusters in the sample. The non-thermal pressure fraction was likely overestimated due to the method used to measure the 3D velocity dispersion. Ideally, the velocity dispersion would only include turbulent velocities, some smaller local bulk velocities were likely included, increasing the total measured velocity dispersion. In contrast, the region used to obtain the bulk velocities in the 2D case was smaller, as each azimuthal bin had its own bulk velocity and so the overall velocity dispersion was smaller in the 2D case than in the 3D.

In contrast to the 3D case, $1-b$ for the 2D profiles increases to 0.851 $\pm$ 0.006 after correction. However, the correction has increased the scatter for both 2D and 3D cases. This agrees with \citet{Pearce2020} who also found a significant increase in the bias when using true gas profiles; as with the above results, most of their 3D HSE mass estimates are higher than the true mass ($1-b > 1$). Similarly, their results obtained from using spectroscopic profiles also increase with the $\alpha$ correction, but not as much as our results. Note that the perturbed clusters have particularly large scatter ($\sigma_{b} \approx 0.3$) so are likely what are producing the high ($1-b$) tail in the 3D results.

\subsection{Is mass bias affected by clumping?} 
\label{sec:clump_bias}

Finally, we address the main point of this section, whether the amount of clumping in the ICM directly affects the hydrostatic mass bias for the 2D case. Since cool, denser clumps of gas increase the clumping value, they should also increase the measured brightness of that gas clump due to the $\rho^2$ dependence within the emission measure, altering the extracted gas density profiles. To account for this effect, we have used gas profiles determined using the azimuthal median. In addition, gas clumps are direct evidence of a non-hydrostatic ICM as it shows that the gas is not evenly distributed, and therefore one would expect that the clumping correlates with the hydrostatic mass bias; it is this effect that we are interested in here.

\begin{table*}
    \centering
    \caption{The median and standard deviation of the bias values when the cluster sample is split into high and low clumping, azimuthal scatters and non-thermal pressure fraction respectively. In each case, the high and low subsamples contain approximately the same number of objects.}
    \begin{tabular}{|l|l|l|l|l|}
    \hline
        ~ & \multicolumn{2}{|c|}{High} & \multicolumn{2}{|c|}{Low}\\ 
        ~ & $\langle 1-b \rangle$ & $\sigma_{1-b}$ & $\langle 1 -b \rangle$ & $\sigma_{1-b}$ \\ \hline
        $\mathscr{C}$        & 0.709 $\pm$ 0.008 & 0.182 $\pm$ 0.008 & 0.792 $\pm$ 0.006 & 0.153 $\pm$ 0.008\\ 
        $\sigma_{A,EM}$      & 0.724 $\pm$ 0.006 & 0.190 $\pm$ 0.008 & 0.779 $\pm$ 0.007 & 0.147 $\pm$ 0.007\\ 
        $\sigma_{A, T_{SL}}$ & 0.725 $\pm$ 0.007 & 0.192 $\pm$ 0.008 & 0.773 $\pm$ 0.007 & 0.146 $\pm$ 0.007 \\ 
        $\alpha$             & 0.734 $\pm$ 0.009 & 0.172 $\pm$ 0.008 & 0.768 $\pm$ 0.009 & 0.169 $\pm$ 0.008 \\ \hline
        \multicolumn{5}{|c|}{Non-thermal pressure corrected}  \\
        \hline
        $\mathscr{C}$        & 0.83  $\pm$ 0.01 & 0.32  $\pm$ 0.01 & 0.854 $\pm$ 0.006 & 0.18 $\pm$ 0.01\\ 
        $\sigma_{A,EM}$      & 0.82  $\pm$ 0.01 & 0.32  $\pm$ 0.01 & 0.867 $\pm$ 0.008 & 0.19 $\pm$ 0.01\\ 
        $\sigma_{A, T_{SL}}$ & 0.86  $\pm$ 0.01 & 0.31  $\pm$ 0.01 & 0.848 $\pm$ 0.006 & 0.20 $\pm$ 0.01 \\ 
        $\alpha$             & 0.88  $\pm$ 0.01 & 0.28  $\pm$ 0.01 & 0.823 $\pm$ 0.009 & 0.23 $\pm$ 0.01\\
        \hline
    \end{tabular}
    \label{table:bias_obs}
    %code: HSE_bias_obs.py
\end{table*}

When the cluster sample is split into two groups of approximately equal size of high and low clumping values at $R_{500}$, it is found that the clusters with lower clumping have smaller bias and scatter (see leftmost panel of Fig. \ref{fig:bias_pdfs} and Table \ref{table:bias_obs}). However, when comparing the clumping directly to the mass bias for individual clusters, only a weak correlation is found, similar to \citet{Ansarifard2020}. 

In addition to the clumping, Fig. \ref{fig:bias_pdfs} also shows the probability distribution function of the mass bias when split into high and low azimuthal scatter in the emission measure (second panel), spectroscopic-like temperature (third panel) and the non-thermal pressure fraction (rightmost panel). The medians and standard deviations of these distributions are shown in Table \ref{table:bias_obs}. 

We find that lower azimuthal scatters and a lower non-thermal pressure fraction also give lower mass biases than the high sub-samples, as well as slightly reducing the scatter in the distribution of bias values.

Fig. \ref{fig:corr_bias_pdfs} shows the same as Fig. \ref{fig:bias_pdfs} but with non-thermal pressure corrected masses. The discrepancy between the high and low mass biases for the clumping and azimuthal scatter in the temperature is essentially eliminated after application of the correction. However, the scatter in the mass bias is still smaller in the low clumping and azimuthal scatter in the emission measure sub samples.

\section{Summary \& Conclusions}
\label{sec:conclusion}
In this paper, we used the results of the MACSIS simulations to investigate the effect of gas inhomogeneities in the ICM of simulated clusters, and how well this correlates with potential X-ray observables, such as the azimuthal scatter. The hydrostatic mass estimate was calculated using both mass weighted 3D profiles and X-ray weighted 2D projected profiles and how this varied with the clumping and projection was also explored. The addition of a mass correction that accounts for the fraction of non-thermal pressure in a cluster was also investigated, resulting in an improvement for the mass estimates calculated using projected profiles.

Our main results can be summarised as follows:

\begin{itemize}
    \item Using the four classification criteria outlined in Section \ref{sec:morph}, we find that approximately one third of clusters are relaxed at $z=0$ and one fifth at $z=0.5$ (Fig. \ref{fig:mass_dist_class}).
    
    \item In agreement with previous work \citep[e.g.][]{Nagai2011GASCLUSTERS, Roncarelli2013, Eckert2015, Planelles2017, Ansarifard2020}, we found that the clumping increases with radius, due to an increasing number of substructures at large radii. In MACSIS clusters, the clumping level remains approximately consistent out to $z=0.5$. Both more massive and more disturbed clusters are found to be clumpier (see top row of Fig. \ref{fig:clump_obs_profs}).
    
    \item As the clumping is a purely theoretical quantity and is not measurable in observations, we investigate potential observational proxies such as the azimuthal scatter in the emission measure and temperature. The level of azimuthal scatter of emission measure in clusters follows that of clumping. It increases with radius, again due to the large number of extended substructures, is stronger in disturbed and massive clusters, and is independent of redshift (see second row of Fig. \ref{fig:clump_obs_profs}).
    
    \item Weightings attempting to match observations were used to obtain the gas property profiles. This caused a deviation from the theoretical mass-weighted property profiles, matching that of \citet{Rasia2005, Roncarelli2018MeasuringObservations}. The spectroscopic-like temperature was $\approx 20$ per cent lower than the mass-weighted temperature at $R_{500}$, resulting in a higher non-thermal pressure fraction when using "observational" weightings rather than mass weightings.
    
    \item The azimuthal median to obtain gas property profiles mitigated the effect of substructures from the profiles. This was particularly evident in the emission measure profiles (see Fig. \ref{fig:mean_median_20}).
    
    \item Using unprojected 3D gas profiles we calculated a median hydrostatic mass bias of $1-b=0.86$ and for the projected 2D profiles we obtained $1-b=0.75$. The difference in these values is a result of both projection effects and the use of weightings to match the spectroscopic profiles. The hydrostatic mass bias was corrected using the non-thermal pressure fraction which resulted in a shift of the mean bias to $1-b=1.15$ for the 3D case and 0.85 for the 2D case. However, despite an improvement in the mass bias for the projected case, the overall scatter increases significantly.
    
    \item We find that clusters with a low clumping, non-thermal pressure fraction and azimuthal scatter (in both the emission measure and temperature) all have a reduced mass bias with a narrower distribution (see Fig. \ref{fig:bias_pdfs}). Clusters with these properties are more likely to be relaxed; there are not any major substructures or recent mergers affecting the extracted gas property profiles. The assumptions of hydrostatic equilibrium and spherical symmetry are therefore more realistic in this case, leading to a smaller bias. 
    
    \item When applying the non-thermal pressure correction to the cluster mass estimates, the discrepancy in the mass bias between clusters with a high and low clumping and azimuthal scatters is reduced (see Fig. \ref{fig:corr_bias_pdfs}). However, the narrower distributions in the more relaxed subsamples remains. The exception to this is in the clusters with a high non-thermal pressure fraction, which is found to now be closer to the true mass than clusters with a low non-thermal pressure fraction.
\end{itemize}

In conclusion, using projected observationally weighted gas profiles increases the hydrostatic mass bias in comparison to using the mass weighted theoretical 3D profiles. Future observations, such as Athena \citep{Nandra2013TheMission}, will be able to take detailed velocity measurements of cluster gas, allowing an accurate estimate for a cluster's non-thermal pressure fraction and its effect on the hydrostatic mass to be studied. However, when a non-thermal pressure fraction correction is used, the cluster-averaged bias is reduced, but the scatter between individual clusters increases for the projected case. This scatter is primarily driven by the morphologically disturbed systems and is most easily reduced by carefully selecting relaxed systems (which may introduce additional biases) via their dynamical state or by selecting clusters with lower azimuthal scatter. This will be key as a well known high bias with a narrow scatter is a more useful result from observations than a lower bias with a larger scatter.

\section*{Acknowledgements}
The authors would like to thank the referee for helpful comments that improved the clarity and quality of this work. This work used the DiRAC@Durham facility managed by the Institute for Computational Cosmology on behalf of the STFC DiRAC HPC Facility (www.dirac.ac.uk). The equipment was funded by BEIS capital funding via STFC capital grants ST/K00042X/1, ST/P002293/1, ST/R002371/1 and ST/S002502/1, Durham University and STFC operations grant ST/R000832/1. DiRAC is part of the National e-Infrastructure.

We also wish to thank the Science and Technologies Facilities Council for providing studentship support for IT. EA acknowledges the STFC studentship grant ST/T506291/1.

\section*{Data Availability}
The data used in the production of this article can be shared upon reasonable request.


\bibliographystyle{mnras}
\bibliography{references} 

% Don't change these lines
\bsp	% typesetting comment
\label{lastpage}
\end{document}